\documentclass[]{interact}

\usepackage{epstopdf}% To incorporate .eps illustrations using PDFLaTeX, etc.
\usepackage{subfigure}% Support for small, `sub' figures and tables

\usepackage{natbib}% Citation support using natbib.sty
\bibpunct[, ]{(}{)}{;}{a}{}{,}% Citation support using natbib.sty
\theoremstyle{plain}% Theorem-like structures provided by amsthm.sty
\newtheorem{theorem}{Theorem}[section]

\newtheorem{corollary}[theorem]{Corollary}

\theoremstyle{definition}
\newtheorem{definition}[theorem]{Definition}

\theoremstyle{remark}

\usepackage{hyperref}
\usepackage[utf8]{inputenc}

\usepackage{setspace}
\usepackage{amsthm, amsmath, amssymb, natbib, mathtools}
\newtheorem{result}{Result}[section]

\newcommand{\argmin}{\argmin{\rm arg \, min}}

\usepackage{xcolor}
\definecolor{dured}{RGB}{144, 0, 0}

\begin{document}

\articletype{}

\title{Modeling Sums of Exchangeable Binary Variables}

\author{\name{Ryan Elmore$^{}$}
\affil{$^{}$University of Denver\\
Daniels College of Business\\
Department of Business Information and Analytics\\
2101 S. University Blvd, Suite 580\\
Denver, CO 80210}
}

\thanks{CONTACT Ryan Elmore. Email: \href{mailto:Ryan.Elmore@du.edu}{\nolinkurl{Ryan.Elmore@du.edu}}}

\maketitle

\begin{abstract}
We introduce a new model for sums of exchangeable binary random variables.
The proposed distribution is an approximation to the exact distributional
form, and relies on the theory of completely monotone functions and the
Laplace transform of a gamma distribution function. Using Monte Carlo
methods, we show that this new model compares favorably to the beta-binomial
model with respect to estimating the success probability of the Bernoulli
trials and the correlation between any two variables in the exchangeable set.
We apply the new methodology to two classic data sets and the results are
summarized.
\end{abstract}


\hypertarget{introduction}{%
\section{Introduction}\label{introduction}}

Correlated binary outcomes, either by design or through natural conditions, is
a common occurrence in many fields. Examples include measuring a Bernoulli
outcome in a repeated measures study, teratological risk assessment, studies of
familial diseases and genetic traits, and group randomization studies, among
many others. Of immediate concern around the world, COVID-19 positivity tests
within a family or geographical unit introduce potentially correlated binary
outcomes.

\citet{kuk:2004} provides a nice introduction to developmental toxicity
studies and the statistical issues therein. We will summarize the details from
their paper as its development is closely related to what is presented here. In
a standard developmental toxicology study, pregnant laboratory animals are
often randomly assigned to receive varying dose levels of a toxic substance
during a major period of organogenesis. Their lives are usually terminated
before giving birth, their uterus is subsequently removed and examined for
possible birth defects. For each litter in such a study, there is a sequence of
Bernoulli random variables \(X_1, X_2,\ldots,X_m\) where \(X_i=0,1\), denoting the
absence or presence of the birth defect.

It is commonly assumed that members of the same litter will behave more
similarly than nonlittermates and, therefore, one may assume a degree of
correlation between littermates. \citet{kuk:2004} notes that litter effect can be
accounted for by assuming the intralitter correlation is induced by a random
effect that is shared by all fetuses in the same litter. This random effect
accounts for all of the environmental and genetic factors that littermates
share in common. \citet{pang:2005} point out that, ``failure to account for litter
effect and the overdispersion it induces will lead to estimates with overstated
precision.''

In earlier work, \citet{will:1975} states that it is necessary to model variation
between fetuses in the same litter and variation between litters receiving the
same treatment. It is this insight that leads to the development of the
beta binomial model for application to toxicological experiments involving
reproduction and teratogenicity. \citet{will:1975} essentially assumes that the
probability of response varies as a beta distribution between dose groups to
model the overdispersion due to litter effect. It should be noted that
\citet{Skel:1948} was first to propose the idea of using the beta distribution to
describe variation in the probability parameter of the binomial distribution.
For a large portion of the past 40+ years, the beta binomial distribution has
been the gold standard when it comes to modeling clustered binary data.
Additional models include a correlated binomial model proposed by \citet{kup_hase},
a correlated beta binomial model discussed by \citet{paul} and \citet{pack}, an extended
beta binomial model introduced by \citet{pren:1986}, and ``additive'' and
``multiplicative'' generalizations of the binomial given in \citet{Alth:1978}.

\citet{Geor:Bowm:1995} developed an exact distribution for sums of exchangeable
binary variables. In addition, they propose an approximating model for
\(\lambda_k=P(X_1=X_2=\cdots =X_k=1)\), \(k=1,2,\ldots,m\) using
\(\lambda_k=f(k;\beta)\) where \(f\) is the completely monotone folded-logistic
function. A drawback to this model is that with only a single parameter \(\beta\),
this model lacks the flexibility of many two-parameter models, such as the
beta binomial model, when estimating success probability and intra-cluster
correlation.

\citet{kuk:2004} notes that the shape of the beta binomial
probability function is often U-shaped, J-shaped, or reverse J-shaped instead of
unimodal with the mode near the expected value of \(mp\). Hence, all of the
probability mass could be concentrated at 0 and \(m\), and a value near the
``expected'' value could be very unlikely. Essentially what happens is that the beta binomial, and other existing distributions, tend to underestimate the risk
of at least one littermate having a birth defect. \citet{kuk:2004} introduced the
\(q\)-power model that is based on the exchangeable theory developed in
\citet{Geor:Bowm:1995}. The shared response model, introduced in \citet{pang:2005}, can model
the data without overestimating the probability of no affected fetuses.

In this manuscript, we propose a new (approximate) distribution for handling
exchangeable binary data. Our model is based on the theory of \citet{Geor:Bowm:1995},
and is similar to \citet{kuk:2004}, \citet{yu2008sums}, and \citet{bowman2016statistical} in its
development. Background information related to several previous models, and our
new model are introduced in Section \protect\hyperlink{model-development}{2}. In Section
\href{Monte\%20Carlo\%20Simulation}{3}, we present the
results of a large-scale Monte Carlo study designed to assess several
statistical properties of the proposed model relative to those of existing
methods. Two classic examples are analyzed in Section \href{Examples}{4} and our concluding remarks are reported in Section \href{Discussion}{5}.

\hypertarget{model-development}{%
\section{Model Development}\label{model-development}}

\hypertarget{prior-research}{%
\subsection{Prior Research}\label{prior-research}}

Modeling finite sums of exchangeable binary random variables is
explored in detail in the papers of \citet{Geor:Bowm:1995}, \citet{Bowm:Geor:1995}, and
\citet{Geor:Kode:1996}, among other papers outlined in Section 1. An exact distribution
for the sum of exchangeable binary random variables is derived in
\citet{Geor:Bowm:1995}. We summarize their results in order to develop our proposed
model.

Let \(\bm{Y}=(Y_1,Y_2,\dots,Y_m)^T\) denote a vector of exchangeable binary random
variables. By exchangeable, we mean that
\begin{equation*}
  (Y_1, Y_2, \dots, Y_k)^T \stackrel{D}{=} (Y_{\pi(1)}, Y_{\pi(2)},\dots, Y_{\pi(k)})^T
\end{equation*}
for any permutation \(\pi\) of the integers \(\{1,2,\dots,k\}\). We are interested
in making inferences on the quantity \(S_m = \sum_{j=1}^m Y_j\) for \(k \le m\).
From a straightforward application of the Inclusion-Exclusion principle in
probability, the exact distribution of these sums can be shown to be
\begin{equation}
  \label{eq:sum_dist}
  P[S_m=s] = {m \choose s}\sum_{k=0}^{m-s}(-1)^k{m-s \choose k}p_{s+k},\ \ \mbox{for} \ \ s = 0,1,\dots,m,
\end{equation}
where,
\begin{align}
  \label{eq:sum_constraint}
  p_j & = P[Y_1=1,Y_2=1,\dots,Y_j=1], \ \ j=1,2,\dots,m, \ \mbox{and} \nonumber \\
  1 & = p_0 \ge p_1 \ge p_2 \ge \dots \ge p_m.
\end{align}

Parameter estimates of \(p_1, \dots, p_m\), and hence any \(k^{th}\)-order
correlation can be found in the following way. An observation \(S_m=s\) is simply
an indicator random variable following a multinomial distribution having cell
probabilities given by \(P[S_m=j]\) for \(j=0,1,\dots,m\). Using the inversion
formula
\begin{equation*}
  p_j = \sum_{k=0}^{m-j}\frac{{m-j \choose k}}{{m \choose k}}P[S_m = m-k],
\end{equation*}
we can find the desired estimates. Variance estimates are computed based on the
distributional properties of multinomial probabilities. Rather than estimating
each individual \(p_j\) using a saturated approach, it is
possible to model these parameters using a function which preserves the
constraints given above, namely that
\begin{align}
  \label{eq:constraints}
  1  & =  p_0 \ge p_1 \ge p_2 \ge \dots \ge p_m,\ \ \mbox{and} \nonumber \\
 &  \sum_{k=0}^{m-s}(-1)^k{m-s \choose k}p_{s+k} \ge 0.
\end{align}

\citet{Geor:Bowm:1995} suggest using the folded-logistic function to model the sequences
of probabilities in order to approximate the model defined in Equation
\eqref{eq:sum_dist}. The folded-logistic function is defined by
\begin{equation*}
  p_x(\beta) = \frac2{1+(x+1)^\beta}
\end{equation*}
for \(x\ge0\) and \(\beta >0\). Under this parameterization, Equation
\eqref{eq:sum_dist} becomes
\begin{equation}
    \label{eq:gb_parm}
  P[S_m=s;\beta] = {m \choose s}\sum_{k=0}^{m-s}(-1)^k{m-s \choose k}\frac2{1+(s+k+1)^\beta},
\end{equation}
for \(s = 0,1,\dots,m\). The estimation problem is now reduced to estimating a
single parameter, \(\beta\), rather than estimating the \(m\) individual \(p\)'s.

\citet{kuk:2004} introduces two additional distributions based on the theory of
completely monotone functions. Note that a function \(\varphi\) is completely
monotone if it possesses derivatives \(\varphi^{(n)}\) of all orders and
\((-1)^n\varphi^{(n)}(\lambda) \ge 0\), for \(\lambda >0\). In the first, Kuk models
the sequence of probabilities \(1=p_0 \geq \ldots \geq p_m\) by
\[
\lambda_k=P(X_1=X_2=\ldots=X_k=1)=p^{k^{\gamma}}
\]
where \(k=0,1,\ldots,m\) and \(0\leq p,\gamma \leq 1\). The parameter \(p\) is the
marginal response probability and the parameter \(\gamma\) controls the degree
of association between littermates. A value of \(\gamma=1\) corresponds
to independence between littermates while a value of \(\gamma=0\) corresponds to
complete dependence between littermates. Under this parameterization,
\eqref{eq:sum_dist} can be written as
\begin{equation}
    \label{eq:pp_dens}
    P(S_m=s;p,\gamma)={m \choose s}\sum_{k=0}^{m-s}(-1)^k {m-s \choose k}p^{(s+k)^{\gamma}}.
\end{equation}
Kuk refers to this model as the \(p\)-power distribution.

Kuk mentions that one may also use the same type of power-family model for
\(X'=1-X\). In this case, \(q=1-p=P(X'=1)=P(X=0)\) and, therefore,
\[
\lambda'_k=P(X'_1=X'_2=\cdots=X'_k=1)=P(X_1=X_2=\cdots=X_k=0)=q^{k^{\gamma}}.
\]
This results in the \(q\)-power probability distribution given by
\begin{equation}
    \label{eq:qp_dens}
    P(S_m=s;q,\gamma)=P(S'_m=m-s|q,\gamma)={m \choose s}\sum_{k=0}^s(-1)^k {s \choose k}q^{(m-s+k)^{\gamma}}
\end{equation}
where \(0\leq q,\gamma \leq 1\). Kuk advocates for the use of the \(q\)-power
distribution over the \(p\)-power distribution when modeling overdispersed binary
data.

\hypertarget{laplace-transform-of-the-gamma-lapgam-distribution}{%
\subsection{Laplace Transform of the Gamma (LapGam) Distribution}\label{laplace-transform-of-the-gamma-lapgam-distribution}}

Our development relies on the theory presented in Section 2.1, along
with the following theory on the difference operator \(\Delta\), as given in
\citet{fellerv2}. The difference operator \(\Delta\) is defined on a sequence \(\{c_n\}\)
to be \(\Delta c_n = c_{n+1} - c_n\). If we apply the difference operator to the
new sequence \(\Delta c_n\), we get another sequence
\(\Delta^2c_n = \Delta(\Delta c_n)\). Similarly, the higher-order differences are
defined recursively by \(\Delta^r c_n = \Delta (\Delta^{r-1}c_n)\), where
\(\Delta^1 = \Delta\). It can be shown that the \(r^{th}\)-order difference can be
written as
\begin{equation}
  \label{eq:r_difference}
  \Delta^r c_n = \sum_{k=0}^r{r \choose k}(-1)^{r+k}c_{n+k}.
\end{equation}
This leads to the following definition.
\begin{definition}[Feller V2]
\protect\hypertarget{def:unnamed-chunk-1}{}{\label{def:unnamed-chunk-1} \iffalse (Feller V2) \fi{} } A sequence \(\{c_n\}\) such that \((-1)^r\Delta^rc_\nu \ge 0\) for all combinations \(r,\nu\) is called a completely monotone sequence.
\end{definition}

Applying these results to the sequence \(\{p_n\}\) given above, we see that if
\(\{p_n\}\) is completely monotone, then the constraints defined in
\eqref{eq:constraints} are satisfied. We will model such a sequence using a
\emph{completely monotone function}. Our main result, summarized next, is a direct
application of the following theorem using the gamma distribution function.

\begin{theorem}[Feller V2]
\protect\hypertarget{thm:unnamed-chunk-2}{}{\label{thm:unnamed-chunk-2} \iffalse (Feller V2) \fi{} } A function \(\varphi\) on \((0,\infty)\) is the Laplace transform of a probability distribution F, iff it is completely monotone, and \(\varphi(0) = 1\).
\end{theorem}

\begin{result}
The function $p_x$ defined by
\begin{equation}
\label{eqn:p_x}
    p_x(\alpha,\beta) = \frac1{[1+\beta x]^\alpha}
\end{equation}
for $x \ge 0$ and $\beta,\alpha >0$ is a completely monotone function.
\end{result}

To see this, let \(F\) be the distribution function of a gamma random variable with mean \(\alpha\beta\) and variance \(\alpha\beta^2\), for parameters \(\alpha,\beta >0\). The Laplace transform of \(F\) is given by
\begin{equation}
    \varphi_\lambda(\alpha,\beta)  = \int_0^\infty e^{-\lambda x}\mathrm{d}F(x) = \frac1{(1+\beta\lambda)^\alpha}.
  \end{equation}
Therefore, \(p_x(\alpha,\beta)\) is a completely monotone function by Feller's result given above.

Similarly to the ideas presented in \citet{Geor:Bowm:1995} and \citet{kuk:2004}, we will use
this function as a model for the sequence \(1 = p_0 \ge p_1 \ge \dots \ge p_m\).
Therefore, an approximate distribution of \(S_m\) under this parameterization is
\begin{equation}
  \label{eq:lap_gam}
    P[S_m=s;\alpha,\beta] = {m \choose s}\sum_{k=0}^{m-s}(-1)^k{m-s \choose k}\frac1{[1+\beta(s+k)]^\alpha}
\end{equation}
for \(s=0,1,\dots,m\). We will refer to this distribution as the Laplace
transform of the gamma distribution, or the LapGam for short.

We wish to emphasize that the distribution defined in
Equation (10) is constructed from the gamma distribution function and the
theory of completely monotone functions. There are no doubt additional
distributions similar to the LapGam that can be defined using distribution
functions characterized by multiple parameters. Hopefully the theory presented
here will spur additional work in the area.

\hypertarget{estimation}{%
\subsection{Estimation}\label{estimation}}

Let \(S_{m1},S_{m2},\dots,S_{mn}\) be a random sample of sums of exchangeable
binary random variables defined by \(S_{mi} = \sum_{j=1}^mY_{ij}\) where
\(P[Y_{ij}=1]=p_1\). We will assume that \(S_{mi}\) follows the distribution given
by (\ref{eq:sum_dist}) with
\begin{equation}
  \label{eq:lap_gam_p}
  p_{s+k} = p_{s+k}(\alpha,\beta) = \frac1{\left[1+\beta(s+k)\right]^{\alpha}}.
\end{equation}
Thus, the log-likelihood function for this sample can be written as
\begin{align}
  \label{eq:exch_ll}
  l_n(\alpha,\beta;\bm{s}) &= \log \left(\prod_{i=1}^nP[S_m=s_i;\alpha,\beta]\right) \nonumber \\
  &= \sum_{i=1}^n \log{m \choose s_i} + \sum_{i=1}^n \log \left(\sum_{k=0}^{m-s_i}(-1)^k{m-s_i \choose k}p_{s_i+k}(\alpha,\beta)\right) \nonumber \\
  &= \sum_{i=1}^n \log{m \choose s_i} + \sum_{i=1}^n \log \left(\sum_{k=0}^{m-s_i} \frac{(-1)^k{m-s_i \choose k}}{\left[1+\beta(s+k)\right]^{\alpha}}\right).
\end{align}
Our interest is in finding maximum likelihood estimators (MLEs)
\(\hat{\alpha}\) and \(\hat{\beta}\). It is
straightforward to write a Newton-Raphson algorithm to maximize the likelihood
given in Equation \eqref{eq:exch_ll}, or simply use an optimization method in R
or python to find the MLEs. We have found that either approach is numerically
stable in optimizing several different variants of this likelihood, for example,
as when using the semiparametric approach described in Section 4.2. The delta
method \citep{lehmann} can then be utilized to find estimators of
the probability parameters and correlations (of potentially all orders).

\hypertarget{monte-carlo-simulation}{%
\section{Monte Carlo Simulation}\label{monte-carlo-simulation}}

In order to assess the performance of the LapGam model in a controlled
environment, we conducted a large-scale Monte Carlo study. Three additional
models were chosen as a basis of comparison in this simulation study: the
\citet{Geor:Bowm:1995} model given in equation \eqref{eq:gb_parm}, the beta binomial
model \eqref{eq:beta_bin_pren} using the parameterization defined in
\citet{pren:1986}, and the \(q\)-power model \eqref{eq:qp_dens} defined in \citet{kuk:2004}.
The models were evaluated in terms of estimating the binary response
probability \(p\) and first-order correlation \(\rho\) for sums of correlated
binary variables. We considered 40 different scenarios corresponding to
\(p=0.1,0.2,0.3,0.4,\) and \(0.5\) and \(\rho=0.05,0.10,0.15,\) and \(0.20\). We
varied the number of Bernoulli trials using \(m=10\) and \(m = 15\). For every
scenario, \(B=1000\) samples of size 100 were simulated. The data for this
simulation study were generated using the \texttt{bindata()} package
\citep{Leisch1} available in the R software \citep{R_proj}. For the sake of brevity, we
only discuss the results of the simulations when \(m = 10\) in this manuscript.
The \(m= 15\) scenario is similar to what is shown here. A comprehensive summary
of the full set of simulation results is available from the author upon
request.

\begin{figure}

{\centering \includegraphics[width=0.9\linewidth]{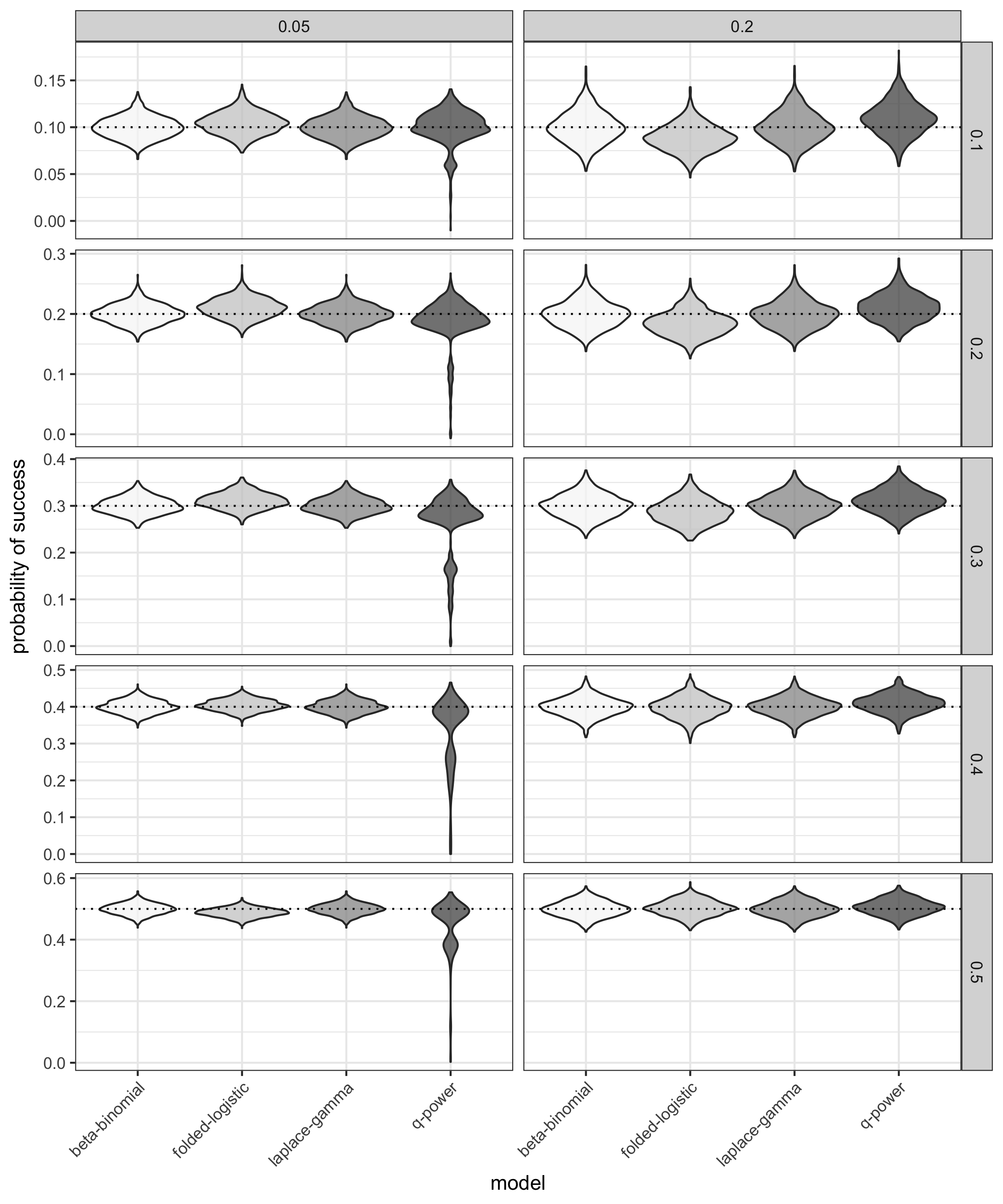} 

}

\caption{The estimated sampling distributions of $\hat{p}$ using the beta binomial, folded-logistic, LapGam, and $q$-power models. Each row and column correspond to particular values of $p$ and $\rho$, respectively, that were used to generate the data in each simulation. The horizontal dashed lines serve as references for the targeted values of $p$ to be estimated.}\label{fig:sim1}
\end{figure}

We first discuss the results of estimating \(p\), the success probability. The
simulated sampling distributions of \(\hat{p}\) can be see in Figure
\ref{fig:sim1}. Specifically, each row (\(p\)) and column (\(\rho\)) combination
corresponds to the parameter values that were used to generate the data. Each
box shows the estimated sampling distributions of \(\hat{p}\) under the four
models in question. As can be seen in this figure, the beta binomial and the
LapGam models perform almost identically across the ten different scenarios
presented here. On the other hand, the estimates of \(p\) based on the
folded-logistic and \(q\)-power models show evidence of bias in certain
situations. For example, there is noticeable bias in both when the
intra-cluster correlation is high (0.2) and the success probability is low
(0.1).

\begin{figure}

{\centering \includegraphics[width=0.9\linewidth]{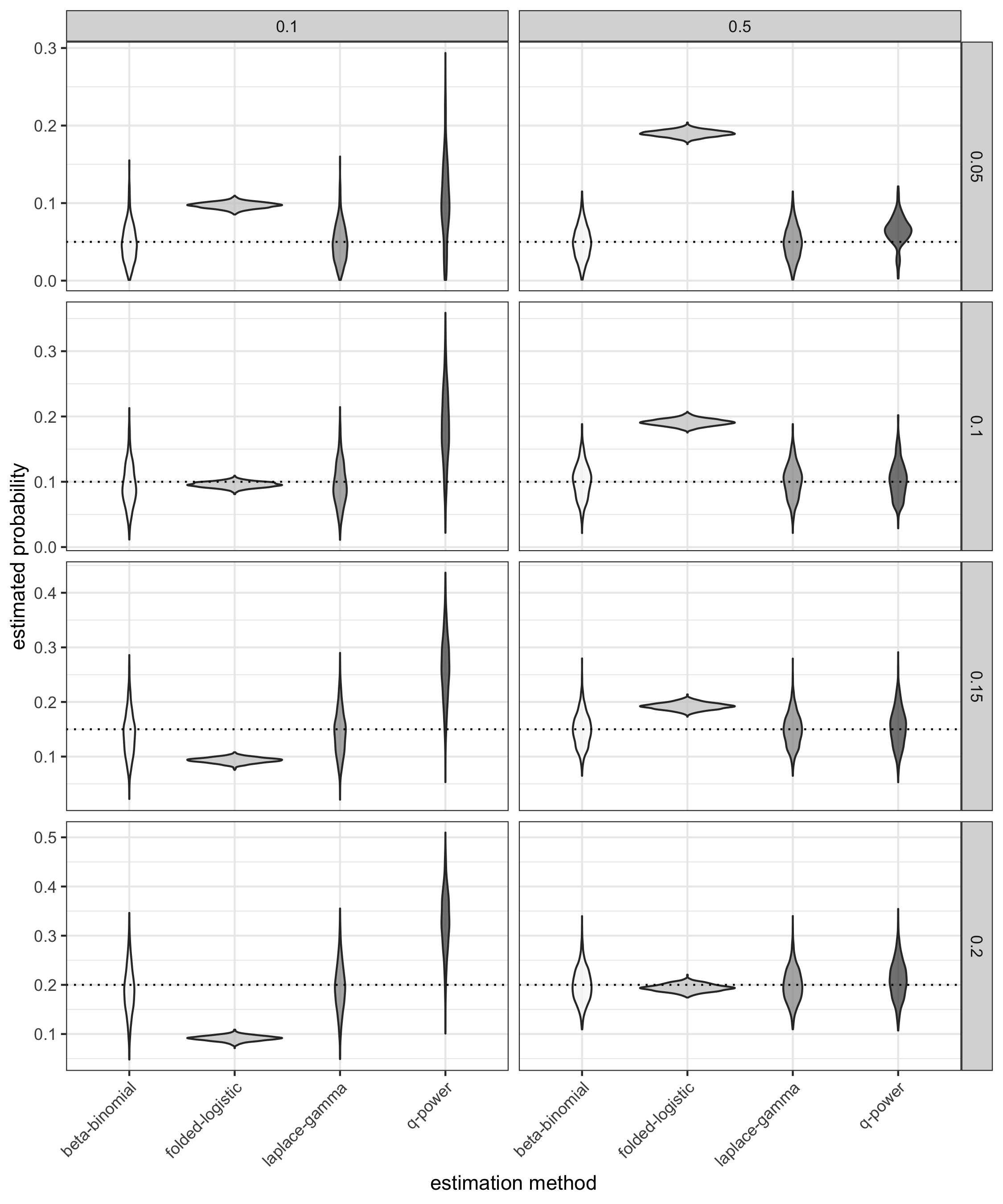} 

}

\caption{The estimated sampling distributions of $\hat{\rho}$ using the beta binomial, folded-logistic, LapGam, and $q$-power models. Each row and column correspond to particular values of $\rho$ and $p$, respectively, that were used to generate the data in each simulation. The horizontal dashed lines serve as references for the targeted values of $\rho$ to be estimated.}\label{fig:sim2}
\end{figure}

Figure \ref{fig:sim2} shows the results when estimating \(\rho\). Each row
(\(\rho\)) and column (\(p\)) combination indicates the parameter values that
were used to generate the data and shows the estimated sampling distributions
of \(\hat{\rho}\) under the four models in question. Similar to the story told
above, the beta binomial and LapGam models tend to perform well at estimating
\(\rho\) in each scenario, whereas the other two show some bias. The
folded-logistic model, in particular, does a bad job at estimating \(\rho\)
when the success probability is 0.5 and the individual trials are weakly
correlated, \(\rho = 0.05\).

\hypertarget{examples}{%
\section{Examples}\label{examples}}

\hypertarget{brassica-data}{%
\subsection{Brassica Data}\label{brassica-data}}

The following example consists of data presented in \citet{Skel:1948} and \citet{Alth:1978} on the secondary association of chromosomes in \emph{Brassica}, a group of plants belonging to the Mustard family (botany.com). If the probability of association is constant within and across nuclei and the individual bivalents are independent, then the counts can be assumed to follow a binomial distribution. However, \citet{Skel:1948} and \citet{Alth:1978} discuss the fact that these data are overdispersed relative to a binomial model and that the beta binomial (and other models) provide an adequate fit. Table \ref{tab:brassica} provides a summary of the data and the expected counts under five models: binomial, beta binomial, folded-logistic, LGa, and the \(q\)-power model.

The \(p\)-values for the chi-square goodness of fit statistics for the beta binomial, LGa, and the \(q\)-power models are 0.8594, 0.8621, and 0.9993, respectively. On the other hand, the usual binomial and the folded logistic fits are rejected according to the chi-square test with \(p\)-values equal to 0.0439 and 0.0048, respectively. The probability of association for a given bivalent and the correlation among pairs of bivalents are given in Table \ref{tab:parms}. As can be seen in this classic example, it is difficult to distinguish which model provides the best summary of these data between the beta binomial, LGa, and the \(q\)-power models. It is fairly easy to discount the estimates based on the folded-logistic function, however.

\hypertarget{brazil-data}{%
\subsection{Brazil Data}\label{brazil-data}}

Our second example consists of data related to a survey of deaths in children
from a particularly poor region in northeast Brazil. The raw data were reported
in \citet{sastry1997nested} and reproduced in \citet{yu2008sums}. The data consist of 1051
unique families with a total of 2946 children. The number of children in the
families ranged from one to eight, with two being the most frequent. The outcome
of interest is a binary variable indicating childhood mortality. The data are
assumed to be exchangeable as they are overdispersed within families
relative to a binomial model.

As discussed in \citet{yu2008sums}, it appears that the mortality rate and
within-family correlation differ as a function of family size. They account
for this difference using a quadratic function of \(m\), family size, in the
logit of \(\mu\), mortality rate, and the log of \(\gamma\) in the beta binomial
model specification defined in Equation \eqref{eq:beta_bin_pren}. Rather than
requiring strict functional forms, we take a semi-parametric approach to
fitting the beta binomial and LapGam models to the Brazil data. That is, we
model the logit of \(\mu\) and the log of \(\gamma\) using cubic splines in \(m\),
i.e.
\[\textrm{logit}[\mu(m)] = \bm{s}(m)^\prime \bm{\eta}_1\]
and
\[\textrm{log}[\gamma(m)] = \bm{s}(m)^\prime \bm{\eta}_2\]
where \(\bm{s}(m)\) is the cubic-spline basis representation of \(m\) and
\(\bm{\eta}_1\) and \(\bm{\eta}_2\) are vectors of parameters to be estimated
from the data. Similarly, we handle within-family differences in \(\alpha\) and
\(\beta\) from Equation \eqref{eq:lap_gam} using the log, or
\[\textrm{log}[\alpha(m)] = \bm{s}(m)^\prime \bm{\eta}_3\]
and
\[\textrm{log}[\beta(m)] = \bm{s}(m)^\prime \bm{\eta}_4.\]
Again, \(\bm{\eta}_3\) and \(\bm{\eta}_4\) are estimable parameter values and
\(\bm{s}(m)\) is the cubic-spline basis representation of \(m\).

\begin{figure}

{\centering \includegraphics[width=0.48\linewidth]{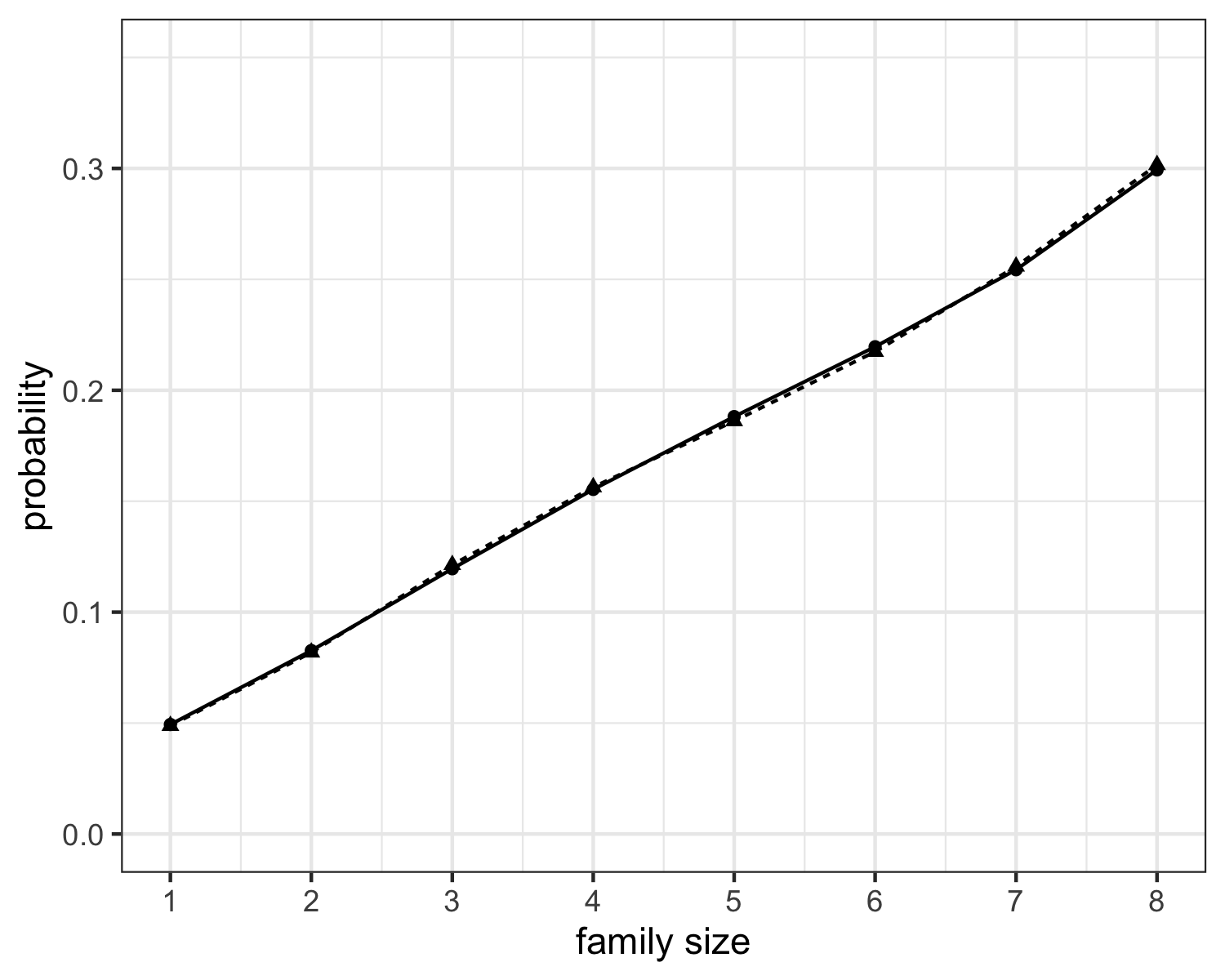} \includegraphics[width=0.48\linewidth]{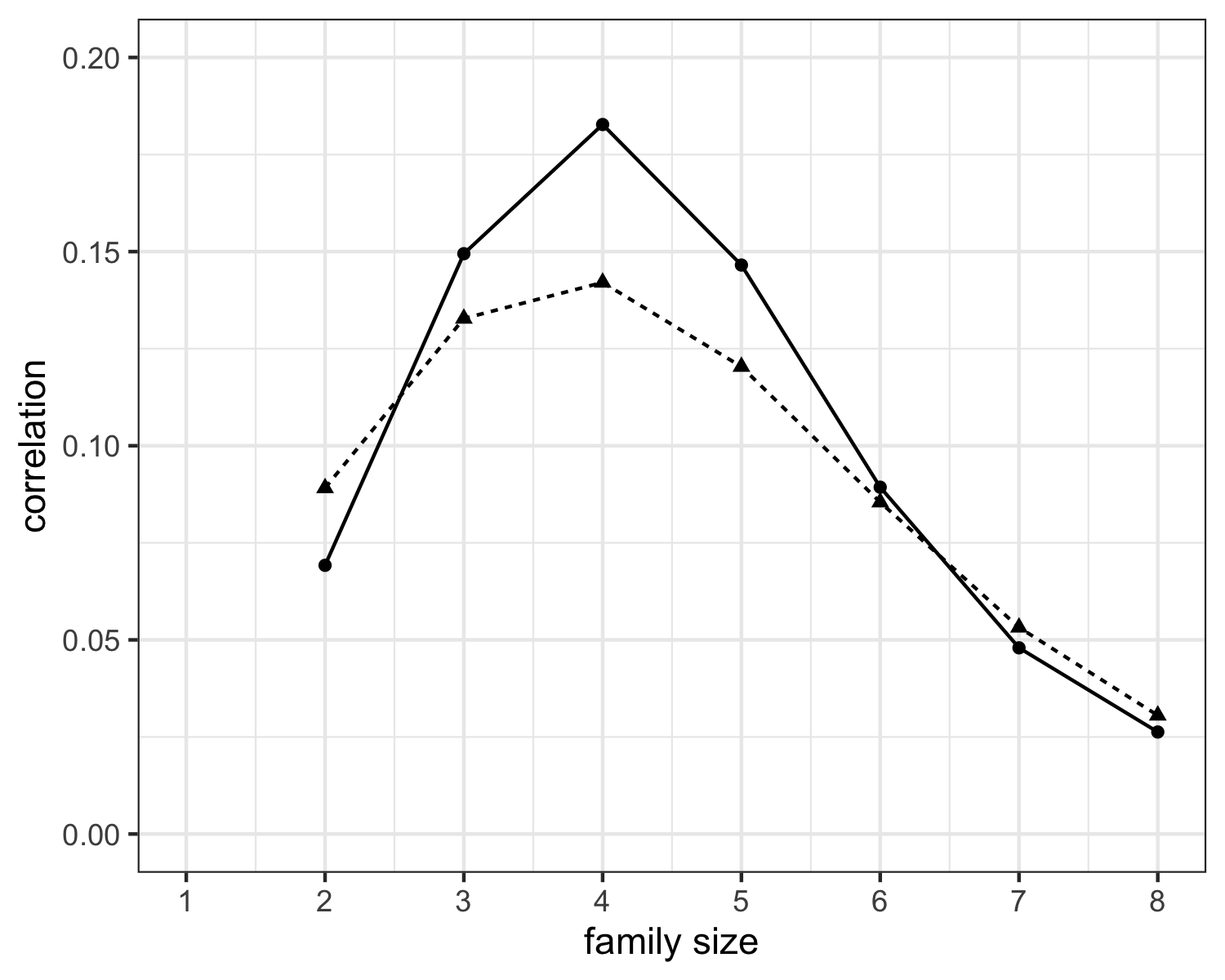} 

}

\caption{The left panel shows the estimated mortality rates as functions of $m$ for the beta binomial model (circles, solid line) and the LapGam model (triangles, dashed). The right panel shows the estimated within-family correlation for each family size ranging from two to eight for the beta binomial model (circles, solid line) and the LapGam model (triangles, dashed).}\label{fig:brazilresults}
\end{figure}

The estimated deviances for the semi-parametric fits of the beta binomial and
LapGam models are 18.79 and 19.09, respectively. The estimated mortality
rates and within-family correlations are presented in Figure
\ref{fig:brazilresults}. The estimates
associated with the beta binomial model are given as circles along the solid
line and the LapGam estimates are shown using triangles on dashed lines. The
estimated probabilities are virtually identical across the two models whereas
there are noticeable differences in within-family correlation. As a comparison
to \citet{yu2008sums} and for completeness, the estimated deviance using a quadratic
function of \(m\) in \(\alpha\) and \(\beta\) is 19.16, suggesting a slightly better
fit when using the semi-parametric model.

\hypertarget{discussion}{%
\section{Discussion}\label{discussion}}

In this paper, we introduced a new model, LapGam, for approximating the
distribution of sums of exchangeable binary variables. The model is
developed using a novel application of completely monotone functions and the
difference operator to the exact distribution as developed in \citet{Geor:Bowm:1995}.
In addition, we demonstrate the efficacy of maximum likelihood estimation of
the LapGam using a large-scale simulation study. Lastly, we demonstrate the use
of this model by applying the results to two classic applications.

The simulation study shows that this new LapGam model performs as well as
the well-known beta binomial distribution under a wide variety of simulated
conditions. These results provide confidence in our conclusions associated
with the two classic examples that we analyzed in Section 4.

As other authors have noted (see \citet{yu2008sums} and references therein), the
exact distribution, as well as its many approximations, rarely out perform
the beta binomial model when applied to estimating parameters in sums of
exchangeable binary variables. This is the case here as well. However, we did
show that our new proposal, the LapGam model, performs as well as the
beta binomial model in all of our simulated scenarios.

In closing, we wish to emphasize that the theory and numerical results that
are presented in this manuscript should serve a building block for future
research on sums of exchangeable Bernoulli variables. That is, we
provide a recipe for the exploration of additional models using the theory of
completely monotone functions and distribution functions, as presented in
\citet{fellerv2}. There are likely similar results to those presented in Section 2.1
that could lead to interesting theoretical and methodological developements.
Hopefully these future studies will come to fruition.

\hypertarget{acknowledgements}{%
\section*{Acknowledgements}\label{acknowledgements}}
\addcontentsline{toc}{section}{Acknowledgements}

The author would like to thank the editorial staff and anonymous referees for
significantly improving the quality of this manuscript.

\begin{table}

\caption{\label{tab:brassica}The observed and expected counts for the {\em Brassica} data using the binomial, folded-logistic, beta binomial, LapGam, and $q$-power models.}
\centering
\begin{tabular}[t]{rrrrrrr}
\toprule
\multicolumn{1}{c}{$m$} & \multicolumn{1}{c}{Observed} & \multicolumn{1}{c}{Binomial} & \multicolumn{1}{c}{FL} & \multicolumn{1}{c}{BB} & \multicolumn{1}{c}{LapGam} & \multicolumn{1}{c}{$q$-power}\\
\midrule
0 & 32 & 24.86 & 50.59 & 33.97 & 33.97 & 32.43\\
1 & 103 & 103.24 & 90.57 & 97.16 & 97.20 & 102.02\\
2 & 122 & 142.93 & 104.45 & 127.67 & 127.61 & 122.69\\
3 & 80 & 65.96 & 91.39 & 78.20 & 78.23 & 79.86\\
\bottomrule
\end{tabular}
\end{table}

\newpage

\begin{table}

\caption{\label{tab:parms}The parameter estimates of $p$ and $\rho$ using the folded-logistic, beta binomial, LapGam, and $q$-power models.}
\centering
\begin{tabular}[t]{lrrrr}
\toprule
\multicolumn{1}{c}{ } & \multicolumn{1}{c}{FL} & \multicolumn{1}{c}{BB} & \multicolumn{1}{c}{LapGam} & \multicolumn{1}{c}{$q$-power}\\
\midrule
$p$ & 0.567 & 0.581 & 0.581 & 0.581\\
$\rho$ & 0.214 & 0.087 & 0.087 & 0.087\\
\bottomrule
\end{tabular}
\end{table}

\hypertarget{appendix}{%
\section*{Appendix}\label{appendix}}
\addcontentsline{toc}{section}{Appendix}

\renewcommand{\thesection}{A}
\setcounter{theorem}{0}

\hypertarget{background-information}{%
\subsection*{Background Information}\label{background-information}}
\addcontentsline{toc}{subsection}{Background Information}

Most of the substantial work on sequences of exchangeable binary variables
revolves around the famous theorem of Bruno de Finetti as stated in \citet{Diac:1977}.

\begin{theorem}[de Finetti]
\protect\hypertarget{thm:deFin}{}{\label{thm:deFin} \iffalse (de Finetti) \fi{} } Let \(\{Y_i\}_{i=1}^\infty\) be an infinite sequence of random variables with \(\{Y_i\}_{i=1}^m\) exchangeable for each \(m\); then there is a unique probability measure \(\mu\) on \([0,1]\) such that for each fixed sequence of zeros and ones \(\{e_i\}_{i=1}^m\), we have
\begin{equation*}
    P[Y_1 = e_1, \dots, Y_m = e_m] = \int_0^1p^s(1-p)^{m-s}\mathrm{d}\mu(p)
  \end{equation*}
where \(s=\sum e_i\).
\end{theorem}

Several comments are in order regarding this theorem. Perhaps most important is
the fact that the unique measure \(\mu\) exists \emph{only if} we have an infinite
sequence of exchangeable binary random variables \(\{Y_i\}_{i=1}^\infty\). This
result is known to fail for finite sets, say \(\{Y_i\}_{i=1}^r\), of exchangeable
binary variables. Fortunately, two finite forms of de Finetti's theorem are
developed in \citet{Diac:1977} and are restated here for completeness.

\begin{theorem}[Diaconis (1977)]
\protect\hypertarget{thm:unnamed-chunk-3}{}{\label{thm:unnamed-chunk-3} \iffalse (Diaconis (1977)) \fi{} } Let \(\{Y_i\}_{i=1}^r\) be an exchangeable sequence which can be extended to an exchangeable sequence of length \(k>r\). Then there is a measure \(\mu_k\) on \([0,1]\) such that if \(e_1, e_2, \dots, e_r\) is any sequence of zeros and ones and \(s=\sum_{i=1}^re_i\), then
\begin{equation}
    \label{eq:diac:77}
    \left|P[Y_1=e_1,\dots,Y_r=e_r]-\int_0^1p^s(1-p)^{r-s}\mathrm{d}\mu_k(p)\right| < \frac{c}{k},
  \end{equation}
where \(c\) is a constant that does not depend on the sequence \(e_i\).
\end{theorem}

\begin{corollary}
\protect\hypertarget{cor:deFin2}{}{\label{cor:deFin2} } Let \(\{Y_i\}_{i=1}^\infty\) be an infinite sequence of random variables with \(\{Y_i\}_{i=1}^m\) exchangeable for each \(m\); then there is a unique probability measure \(\mu\) on \([0,1]\) such that for each fixed sequence of zeros and ones \(\{e_i\}_{i=1}^m\), we have
\begin{equation*}
    P[S_m=s] = {m \choose s}\int_0^1p^s(1-p)^{m-s}\mathrm{d}\mu(p)
  \end{equation*}
where \(s=\sum e_i = 0,1,\dots, m\).
\end{corollary}

Results such the previous three are potentially one motivation, but certainly
not the only one, for using the beta binomial distribution to model sums of
correlated binary variables, see \(e.g.\) \citet{Skel:1948}, \citet{grif:1973}, \citet{will:1975},
and \citet{pren:1986}. The beta binomial model is defined on \(S_m\) for sums of
correlated binary random variables having latent response probability \(p\) as
\begin{equation}
  \label{eq:beta_bin_cond}
  P[S_m=s|p] = {m \choose s}p^s(1-p)^{m-s} \ \ \mbox{for}\ s=0,1,\dots,m,
\end{equation}
where \(p \sim Beta(\alpha,\beta)\). This implies that, unconditionally,
\begin{align}
  \label{eq:beta_bin}
  P[S_m=s]  &=  \int {m \choose s}p^{s}(1-p)^{m-s}\mathrm{d}G(p) \nonumber \\
  &=  \frac{{m \choose s}}{\textrm{B}(\alpha,\beta)}\int_0^1p^{\alpha+s-1}(1-p)^{m+\beta-s-1}\mathrm{d}p \nonumber \\
 &= \frac{{m \choose s} \textrm{B}(\alpha+s,m+\beta-s)}{\textrm{B}(\alpha,\beta)}, \ \ \mbox{for} \ s = 0,1,\dots,m
\end{align}
with B\((a,b)=\frac{\Gamma(a)\Gamma(b)}{\Gamma(a+b)}\). The critical, yet non-verifiable assumptions in using such a model as the distribution of finite sums of exchangeable binary variables are that: (i) the set of exchangeable variables can be embedded into an infinite sequence of exchangeable binary random variables, and (ii) the unique measure given in Theorem \ref{thm:deFin}, \(\mu\), is a beta distribution function.

\citet{pren:1986} introduces an alternative form of the beta binomial distribution by defining \(\mu=\alpha(\alpha+\beta)^{-1}\) and \(\gamma=\rho(1-\rho)^{-1}\). The mass function under this parameterization is defined by
\begin{equation}
  \label{eq:beta_bin_pren}
  P[S_m=s;\mu,\gamma] = \frac{\displaystyle {m \choose s} \prod_{a=0}^{s-1}(\mu+\gamma a) \prod_{a=0}^{m-s-1}(1-\mu+\gamma a)}{\displaystyle \prod_{a=0}^{m-1}(1+\gamma a)},
\end{equation}
\(s = 0,1,\dots,m\), for \(0 < \mu < 1\) and \(\gamma > 0\). Using this form, \(\rho=\gamma(1+\gamma)^{-1}\) and small negative correlations are permissible.
In fact, one can show that
\begin{equation*}
  \rho \ge \mbox{max}\{-\mu(m-\mu-1)^{-1},-\bar{\mu}(m-\bar{\mu}-1)^{-1}\},
\end{equation*}
or equivalently,
\begin{equation*}
  \gamma \ge \mbox{max}\{-\mu(m-1)^{-1},\bar{\mu}(m-1)^{-1}\},
\end{equation*}
where \(\bar{\mu}=1-\mu\). Note that usual binomial variation corresponds to
\(\gamma=0\), whereas, binomial variation corresponds to infinite parameter
values under the original parameterization.

\bibliographystyle{tfcad}
\bibliography{bibfile.bib}

\end{document}